\documentclass{elsart}
\usepackage{natbib}
\usepackage{epsfig}

\def\plottwo#1#2{\centering \leavevmode
\epsfxsize=.35\textwidth \epsfbox{#1} \hfil
\epsfxsize=.35\textwidth \epsfbox{#2}}

\newcommand{\apj}{ApJ}
\newcommand{\apjl}{ApJL}
\newcommand{\mnras}{MNRAS}
\newcommand{\aj}{AJ}
\newcommand{\aap}{A\&A}
\newcommand{\pasj}{PASJ}

\def\gta{\mathrel{\rlap{\lower 3pt\hbox{$\sim$}}
        \raise 2.0pt\hbox{$>$}}}

\begin{document}
\runauthor{H.~J.~Buttery, G.~Cotter, R.~W.~Hunstead and E.~M.~Sadler}

\begin{frontmatter}
\title{Searching for Clusters with SUMSS}
\author[cam]{Helen~J.~Buttery}\thanks{HJB would like to thank PPARC for a PhD studentship}
\author[cam]{Garret~Cotter}
\author[syd]{Richard~W.~Hunstead}
\author[syd]{Elaine~M.~Sadler}

\address[cam]{Astrophysics Group, Cavendish Laboratory, Cambridge}
\address[syd]{School of Physics, The University of Sydney, NSW 2006, Australia}

\begin{abstract}

Statistical overdensities of radiosources in the Sydney University
Molonglo Sky Survey (SUMSS) are used as signposts to identify
high-redshift clusters of galaxies. These potential clusters have been
observed at 20 and 13 cm at the Australia Telescope Compact Array
(ATCA) to obtain better positional accuracy for the sources. A
subsample have been imaged in $V$, $R$ and $I$ at the 2.3-m telescope
at Siding Spring and in $J$ and $K$ at the Anglo Australian Telescope
(AAT) and the New Technology Telescope (NTT) at La Silla, Chile. The
colours obtained from these observations will be used to estimate
redshifts for the potential cluster members.

\end{abstract}
\end{frontmatter}

\section{Introduction}

Evidence is growing for a $\Omega_{\Lambda}\sim0.7$, $\Omega_M\sim0.3$
universe (see eg. \cite{deBernardis2002} \cite{Slosar2002}). Structure 
formation theory predicts
that such a universe will contain large numbers of clusters out to
$z\approx1$ \cite{Jenkins2001}.  However, only a handful of such
clusters have been discovered to date (see eg. \cite{Stanford97},
\cite{Rosati99}, \cite{Chapman2000},
\cite{Fabian2001}, \cite{Nakata2001}, \cite{Thompson2001},
\cite{Haines2001}, \cite{Stanford2002}, \cite{Hashimoto2002},
\cite{Benoist2002}); the traditional methods for finding clusters,
using overdensities of optical or IR sources, suffer severely from
contrast and contamination problems at high redshifts. For this reason
the new generation of X-ray satellites---Chandra and XMM-Newton---are
now involved in searches for such high-redshift clusters.

However, determination of the statistical properties of distant
clusters solely on the basis of X-ray surveys may lead to biases: the
density-squared dependence of X-ray bremsstrahlung radiation means
that highly centrally condensed systems (i.e. relaxed and with a well
established cooling flow) will be preferentially selected by
X-ray samples, even though they may not be typical of systems
of a given mass at a given epoch. Other selection techniques are
clearly needed to complement the X-ray work.

Radiosources are widely regarded as tracers of high density regions at
higher redshifts (see eg.  \cite{Venemans2002}). Work on this subject
has been documented by many authors: distant clusters have been found
using deep imaging around single radiosources (see
eg. \cite{Nakata2001}, \cite{Best2000}); also, groupings of
radiosources in the NRAO VLA Sky Survey (NVSS) have proven successful
as tracers of high-$z$ clusters of galaxies (see \cite{Croft2001},
\cite{Cotter2002}). Such techniques are likely to find only a small
subset of the total population and will inevitably involve some
bias. Interactions between galaxies are thought to trigger
radiosources (see eg. \cite{Best2002}). It is therefore likely that by
searching for groupings of radiosources we will bias our search
towards merging clusters. This is complementary to the X-ray searches
which will be biased towards relaxed systems.

\section{Using the SUMSS catalogue to identify potential clusters}

The SUMSS is a radio survey at 0.8 GHz with a resolution of 40 arcsec,
complete to about 6mJy (see \cite{Bock99}) and therefore very similar
in frequency, resolution and completeness to the NVSS.  We have used
the SUMSS catalogue \cite{Mauch2002} to search for overdensities of
radiosources in the Southern Hemisphere.  By searching for groupings
of 5 radiosources within a 7 arcmin radius, we produced a list of 120
candidate clusters. After examining the SuperCOSMOS $R$-band images of
these fields, we were able to remove obvious chance alignments where
several of the sources had clearly unrelated IDs, and some low-$z$
clusters (i.e., a small number of Abell-type clusters). This procedure
left 60 candidates for clusters expected to lie at $z \gta 0.3$,
beyond the SuperCOSMOS $R$-band plate limit for radiosource host
galaxies.

\section{ATCA observations}

\begin{figure}[h!]
\plottwo{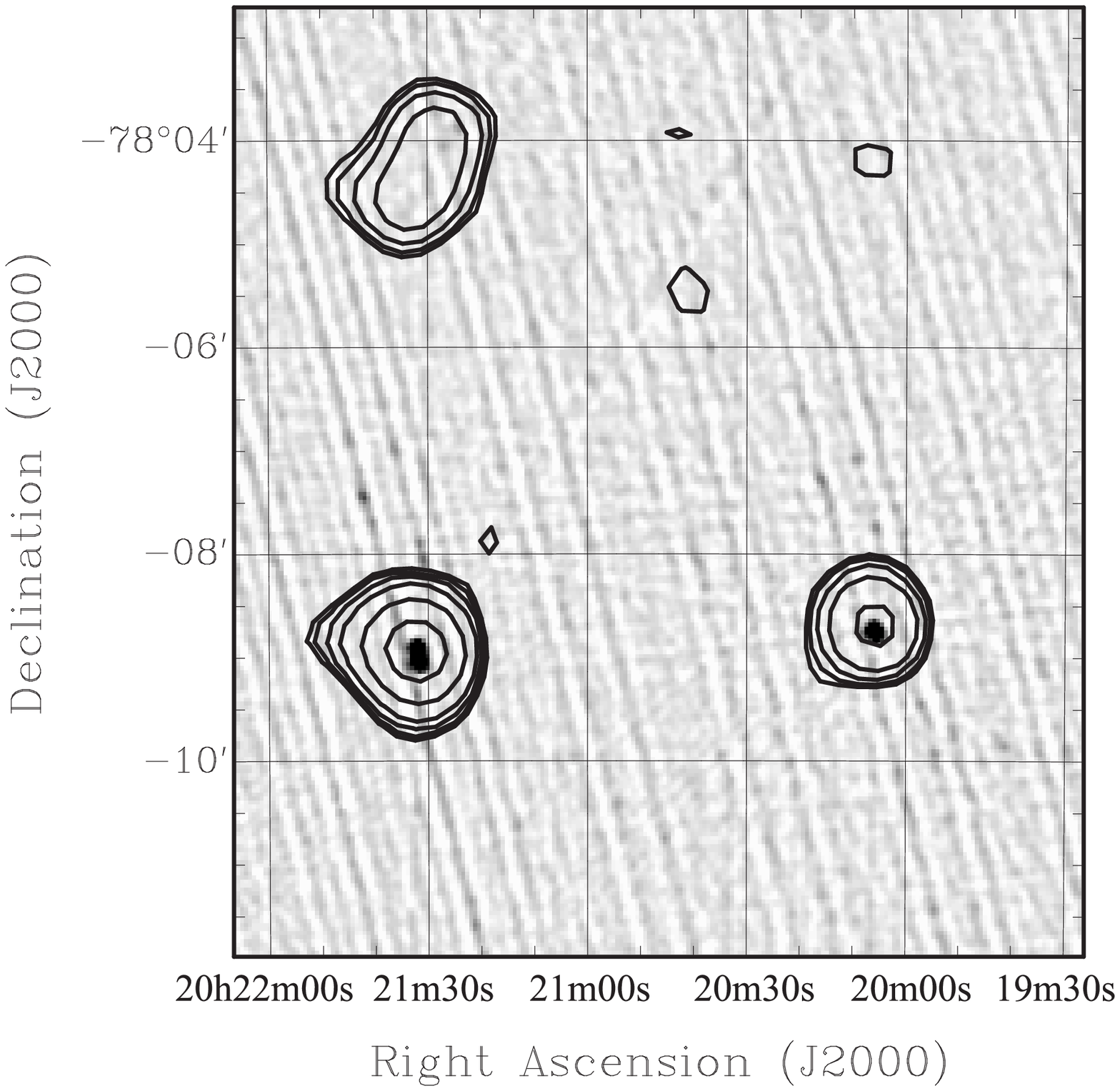}{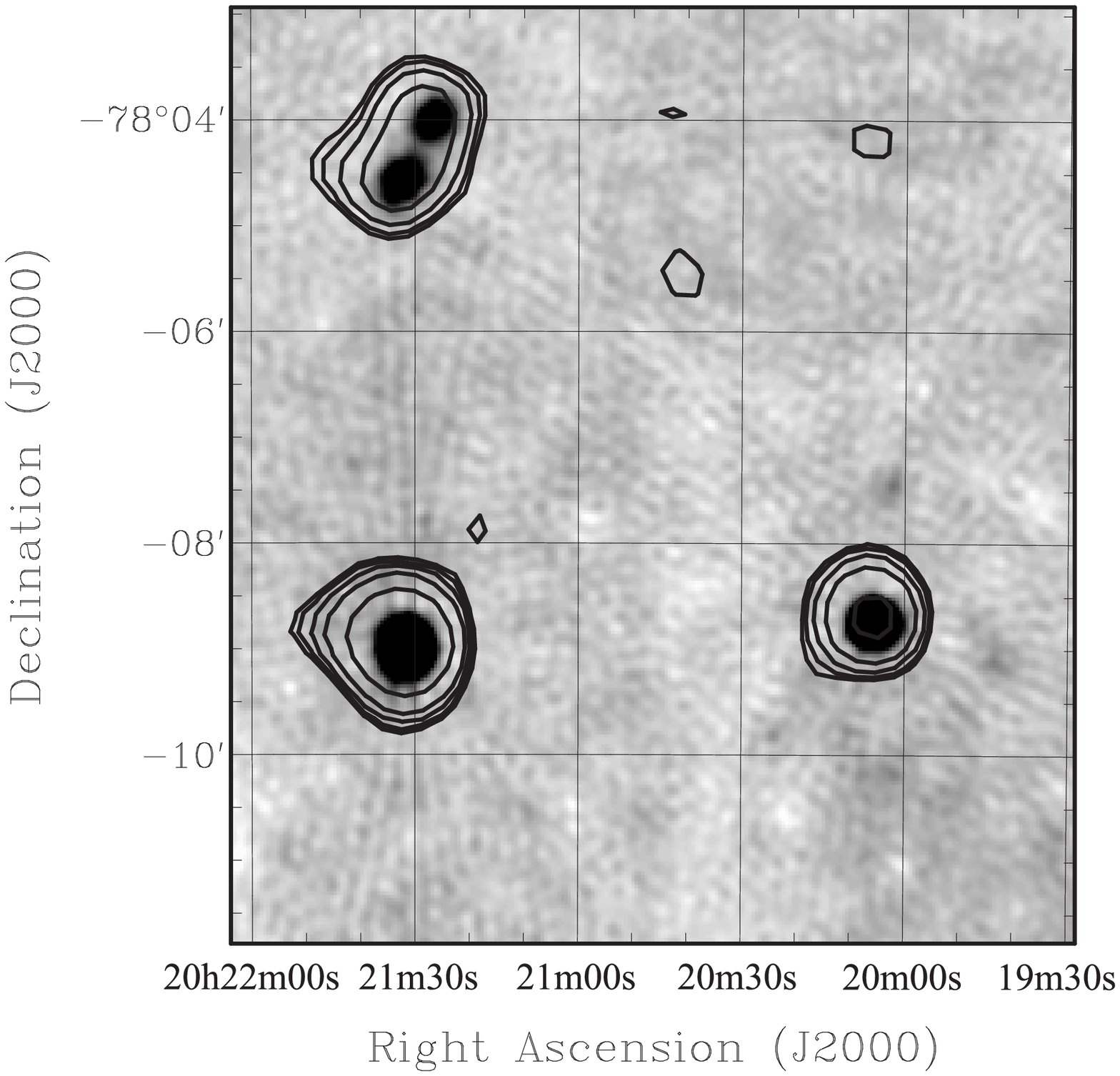}
\caption{The left hand figure shows the ATCA 6km array map at 20cm in greyscale
overlaid with the SUMSS contours at 5, 6, 8, 12, 24 and  48 mJy. 
The top left source is completely resolved 
out. The right hand figure shows the ATCA 1.5km array in greyscale overlaid 
with the same SUMSS contours. The top left source is now apparent - confirming that 
it represents extended emission.
\label{extended}}
\end{figure}

All 60 of these candidates have been further observed at 20cm and 13cm
with the Australia Telescope Compact Array (ATCA) (using the 6-km
array) (see \cite{Buttery2001b}). These observations were carried out
with the goal of providing more accurate source positions to
facilitate optical identifications. However, they have proven very
interesting in their own right.  About thirty percent of the
candidates show diffuse emission, which is often associated with
clusters of galaxies (see \cite{Buttery2002}).  Further observations
of these clusters were made on the ATCA in 2002 November using the
1.5-km array in order to investigate this extended emission (see
Fig. \ref{extended}).

\section{Optical Imaging}

In addition to the radio observations, 40 of the potential clusters
have been imaged to $V \sim 23$, $R \sim 23 $ and $I \sim 22$ with the
ANU 2.3-m telescope. Analysis of this data is ongoing.  These optical
observations are designed to identify clusters out to a redshift of $z
\approx 0.7$, and to make an approximate redshift estimate by locating
the position of the $4000$-\AA~break.

\begin{figure}[h!]
\centering \leavevmode
\epsfxsize=.35\textwidth \epsfbox{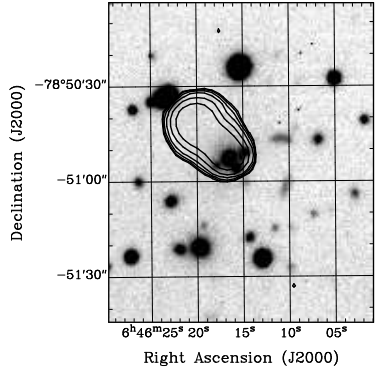}
\caption{The figure shows the ATCA 6km array map at 20cm contours at 0.5, 0.6, 0.8, 1.2, 2.4 and  4.8 mJy overlaid on the $V$-band image from the 2.3-m telescope at Siding Spring. As well as an ID for the radiosource many galaxies of similar magnitude can be seen in the 
image - indicating a low-$z$ cluster in the field. 
\label{optical}}
\end{figure}

Fig. \ref{optical} shows an optical identification for a radiosource
imaged by the ATCA as well as many galaxies of a similar magnitude
around the radiosource.  This cluster has an estimated redshift of
$z\approx0.4$.

The candidates which have no identifications in the optical images are
very likely to contain clusters associated with at least some of the
radiosources. In the FIRST survey (1.4 GHz, $\sim$1mJy completeness)
about $10\%$ of the sources are at a $z>0.7$ (Culverhouse
priv. comm.). SUMSS, which is closely matched in frequency and
completeness, should have a similar redshift distribution, and so the
Poisson probability of five unrelated radiosources lying at different
redshifts $z>0.7$ would be one in $10^5$.

\section{IR imaging}

Service observations have been made of 3 clusters in $J$ and $K$ using
IRIS2 at the Anglo-Australian Telescope. Further imaging in $J$ and
$K$ was carried out using SofI in 2002 December on the New
Technology Telescope in Chile, on the cluster candidates with no
identifications from the 2.3-m. This allows redshift estimation
for clusters where the $4000$-\AA~break lies between $R$ and $J$, or
further to the red.

\begin{figure}[h!]
\centering \leavevmode
\epsfxsize=.35\textwidth \epsfbox{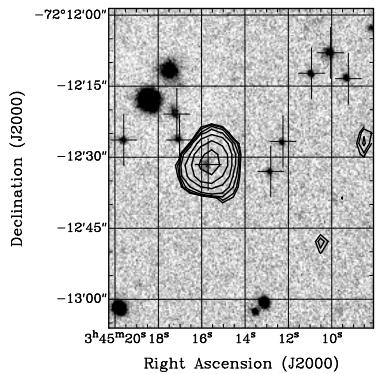}
\caption{The figure shows the $J$-band image taken with the NTT 
showing 
numerous IR sources surrounding one of the radiosources, as well as a 
radiosource ID. The sources labelled with crosses are not present (or only 
very faintly present) in the $R$-band image. This indicates the 
$4000$\AA~break lies between $R$ and $J$, implying a redshift of 1. 
The figure has ATCA
20-cm contours at 0.5, 0.6, 0.8, 1.6, 
3.2, 6.4 and 12 mJy overlaid
\label{IR}}
\end{figure}

Preliminary results show the IR observations have been very succesful
at detecting high-$z$ clusters. Fig. \ref{IR} shows a radiosource
surrounded by $J$-band sources on the left which are not seen in the
$R$-band image. This indicates the $4000$-\AA~break lies between $R$
and $J$ implying a redshift of $z\sim1$.

\section{Conclusions}

Our results give further evidence to indicate that radiosources trace
high-density regions at higher redshifts. Our method of searching for
overdensities of radiosources as a tracer for clusters has been
sucessful.  We now have a sample of several tens of clusters and
robust candidate distant clusters selected by this new technique.  The
clusters will form an ideal sample for future 8-m optical/IR and X-ray
follow-up.  8-m spectroscopy will be essential to confirm the
redshifts of the high-$z$ candidates. X-ray follow-up will allow the
testing of the hypothesis that our technique is biased towards finding
merging systems, by giving information about the morphology of the
clusters. It may also provide new information on the interaction
between radiosources and the cluster environment.

\end{document}